\documentstyle[12pt,epsfig]{article}
\begin{document}


\title{Spin glasses on Bethe Lattices for large coordination number}
\author{ Giorgio Parisi$^{a}$ and Francesca Tria$^{b}$}

\maketitle

\begin{center}

(a) Dipartimento di Fisica, Sezione INFN, SMC and
 UdRm1 of INFM ,\\
Universit\`a di Roma ``La Sapienza'',
Piazzale Aldo Moro 2,
I-00185 Rome (Italy)

b) Dipartimento 
di Scienze Fisiche and Unit\`a INFM, Universit\`a ``Federico II'', \\
Complesso Monte S. Angelo, I--80126 Napoli 
(Italy) 


\end{center}

\vspace{2cm}

\begin{abstract}

\noindent

We study spin glasses on random lattices with finite connectivity.  In the infinite connectivity 
limit they reduce to the Sherrington Kirkpatrick model.  In this paper we investigate the expansion 
around the high connectivity limit.  Within the replica symmetry breaking scheme at two steps, we 
compute the free energy at the first order in the expansion in inverse powers of the average 
connectivity ($z$), both for the fixed connectivity and for the fluctuating connectivity random 
lattices.  It is well known that the coefficient of the $1/z$ correction for the free energy is 
divergent at low temperatures if computed in the one step approximation.  We find that this annoying 
divergence becomes much smaller if computed in the framework of the more accurate two steps 
breaking.  Comparing the temperature dependance of the coefficients of this divergence in the replica 
symmetric, one step and two steps replica symmetry breaking, we conclude that this divergence is an 
artefact due to the use of a finite number of steps of replica symmetry breaking.  The $1/z$ 
expansion is well defined also in the zero temperature limit.

\end{abstract}



\newpage

\section{Introduction}
\noindent Many studies have been devoted to finding analytic solutions of more realistic models 
than the Sherrington-Kirkpatrick one.   The diluted spin glass models belong to this class, and they
are characterized by a finite coordination number;
these models are also interesting because they are connected with different optimization problems 
\cite{sat}.

In the present work we consider lattices where each site is connected with a finite 
number of  randomly chosen sites; we study both the cases where the 
connectivity is fixed and where the connectivity is a Poissonian variable with given 
mean value. The spin interaction is only among nearest neighbour pairs.

The random structure of these lattices allows us to neglect the probability of closed paths of 
finite length: this probability becomes indeed zero in the thermodynamic limit: the correlations 
among the neighbours of a given spin can be neglected.  We are therefore dealing with mean field 
models, although the difficulties due to the finite connectivity don't allow us to solve them exactly.

Whereas in the SK model only the overlap between two replicas occurs as order parameter (the order 
parameter is a function in the infinite step replica symmetry breaking solution and a pure number 
when the replica symmetry is exact), in the finite connectivity models the order parameter becomes a 
function of the overlaps of any number of replicas and then it becomes a function of an infinite 
number of variables when the symmetry is totally broken; as a result it is extremely difficult to 
find the exact free energy \cite{mp86,temp,WonShe}.  In other words the probability distribution of the effective cavity 
fields is Gaussian in the SK model as a consequence of the central limit theorem, so it can be 
characterized by its variance.  When the number of neighbours $z$ is finite, this distribution is no 
more Gaussian and all the moments are relevant and this leads to the presence of an infinite numbers 
of order parameters (also in the replica symmetric situation).

Perturbative solutions have been investigated both near the critical temperature 
\cite{temp} and near the infinite connectivity point (SK model) \cite{primo}, \cite{mio}, 
\cite{ultimo}.  Recently it has been proposed proposed a general non perturbative solution 
developing the Bethe-Peierls cavity method to an approximation that is equivalent to a 
one step replica symmetry breaking level \cite{cavita}.

The present work addresses to the study of the large connectivity expansion: we compute the first 
order of the expansion in the inverse power of the connectivity ($z$) for the free energy.  The 
$1/z$ expansion has been studied for the fixed connectivity model by Goldschmidt and De Dominicis, 
at the first step of replica symmetry breaking \cite{mio}; they found results that exhibit a low 
temperature divergence for the first order correction in $1/z$ to the free energy density.  The 
$1/\sqrt{z}$ expansion, that they computed a $T=0$ and to the second step of replica symmetry 
breaking, has finite corrections both at the first and the second order \cite{ultimo}.  The 
$1/\sqrt{z}$ coefficient becomes yet smaller by a factor ten and by a factor three when one goes 
from the replica symmetric solution to the 1RSB one and from this to the 2RSB one respectively.

In this paper we have computed the coefficient of the $1/z$ expansion up to the second step of 
replica symmetry breaking.  For the replica symmetric and the 1RSB solutions our results agree with 
those found by Goldschmidt and De Dominicis.  These results suggest that the pathological behaviour 
(i.e. the low temperature divergence) is a consequence of the fact that one stops the computation at 
a finite step of the iterative process for breaking the replica symmetry.  Our results indicate that 
the $1/z$ expansion is well defined and can be used also in the zero temperature limit.  We notice 
that a well defined $1/z$ expansion is possible for a model with continuos varying coupling only if 
(irrespective of the sign) the zero temperature entropy  is zero in the limit $z \to \infty$.  Indeed 
it is easy to prove using the approach of \cite{cavita} that in the mean field approximation for 
finite $z$ the zero temperature entropy is identically zero, so that the two limits $z \to \infty$ 
and $T \to 0$ could not be exchanged in an hypothetical model with continuos coupling if the zero 
temperature entropy were different from zero at $z=\infty$.

The paper is organized as follows: in the second section we present the two models we study and the 
high connectivity expansion is obtained.  We use a simple method to evaluate sums over multiple 
replicas overlaps.  In the third section we show how to perform sums over the replica indices in a 
simple way and in section four we illustrate our numerical results for the two steps 
replica symmetry breaking and we compare them with the known ones at the first step of replica 
symmetry breaking. Finally we present our conclusions. The appendix is devoted to a consistency 
check for the form of the free energy we use.

\section{The large connectivity expansion.}

\noindent
By definition  a Bethe lattice is a lattice where the Bethe-Peierls approximation is exact; 
this is equivalent to saying that there are no finite size loops.

In the random lattices we study the typical length of a loop is proportional to 
$\log{N}$: in the infinite volume limit it is therefore a Bethe lattice.  This is locally 
equivalent to a tree-like structure, nevertheless by defining the Bethe lattice as a 
random lattice one bypasses the problem of fixing the boundary conditions to introduce 
frustration (this is provided by the loops of size $ \sim \ln N$).

\subsection{Random lattice with fixed connectivity}
We therefore follow a variational formulation using in the framework of the replica 
approach the same scheme \cite{mio,MPV}.  We define a 
functional and we show that the free energy is obtained as the stationary point with 
respect to an order parameter that will be defined.
The free energy functional is \cite{mio}:
\begin{eqnarray} \label{funz}
n \beta f_n(g_n) & \equiv & z \ln{(Tr_{\{\sigma_a\}}
 g_n^{z + 1}(\{\sigma_a\}))} + \\
 & - & \frac{z + 1}{2} \ln \left\{ \int_{-\infty}^{+\infty}dJ P(J)
Tr_{\{\sigma_a\}}Tr_{\{\tau_a\}} g_n^{z }(\{\sigma_a\})
 g_n^{z }(\{\tau_a\}) \exp{[\beta J \sum_{a = 1}^{n}\sigma_a
\tau_a]}\right\} \nonumber
\end{eqnarray}
where $Tr_{\{\sigma_a\}}$ is the sum over the $2^n$ configurations of the variables $\sigma_a$ with 
$a = 1,\cdots,n$ , $z +1$ is the lattice connectivity, $g_n(\{\sigma_a\})$ is a function of the $n$ 
variables $\sigma_a$ and it plays the role of the order parameter.  Our goal is to make stationary 
the functional $f_n$ with respect to the variation of $g_n(\{\sigma_a\})$.  We have then to find the 
solution of the equation:
 \begin{equation}
\frac{\delta f}{\delta g_n} = 0 \ ,
\end{equation}
that  gives   for the order parameter the equation:
\begin{equation}\label{general}
 g_n(\{\sigma_a\}) = C \int_{-\infty}^{+\infty}dJ P(J)
\sum_{\{ \tau_a \}} 
\exp{\left( 
\sum_{a = 1}^{n} \beta J \sigma_a \tau_a \right)}
g_n^z (\{\tau_a\})\label{EqG}
\end{equation}
with
\begin{equation} \label{cost}
C = \frac{Tr_{\sigma_a} g_n^{z + 1}(\{\sigma_a\})}
{\int_{-\infty}^{+\infty}dJ P(J)
Tr_{\{\sigma_a\}}Tr_{\{\tau_a\}} g_n^{z }(\{\sigma_a\})
 g_n^{z }(\{\tau_a\}) \exp{[\beta J \sum_{a = 1}^{n}\sigma_a
\tau_a]}} \ .
\end{equation}
We can notice that the functional defined by (\ref{funz}) is independent of the $g_n$ 
normalization; we can then use a convenient one, provided one changes the constant $C$ (in 
(\ref{cost})) in $C d^{(1 - z)}$ when changing $g_n$ in $g_nd$.
The correctness of this functional (\ref{funz}) has been proved by De Dominicis et al.  
\cite{mio}; a simple way to get this result is reported for completeness in the appendix.

In order to write the order parameter $g_n(\{\sigma_a\})$ 
in a more explicit form (where the multiple overlaps appear)
we generalize the identity
 \begin{equation} 
\exp{(\beta J \sigma_a \sigma_b)} =
\cosh{(\beta J)} (1 + \sigma_a \sigma_b \tanh{(\beta J))}
\end{equation}
to
\begin{equation} \label{form}
\exp{\left(\beta J \sum_{a=1}^{n} \sigma_a \sigma_b \right)} =
\cosh^n{(\beta J)} \sum_{r = 0}^{n} \left( \tanh^r{(\beta J)}
\sum_{a_1< \cdots <a_r} \sigma_{a_1} \tau_{a_1} \cdots \sigma_{a_r}
 \tau_{a_r} \right) \, ,
\end{equation}
where the last sum is over all possible sets of $r$
replicas, counting once any permutation.

\subsubsection{Interaction with a bimodal distribution.}  
\noindent

We first study the following distribution for the $J$: 
\begin{equation} 
P(J) = \frac{1}{2}[ \delta (J+J_0) + \delta (J-J_0)] \ .   
 \end{equation}
Equation (\ref{form}) is formally identical after  averaging
on the $J$, providing one  sums only over the even $r$ 
and writes  $J_0$ instead of $J$.

If we define the overlaps
\begin{equation}\label{usato} 
q_{a_1 \cdots a_r} = \frac{ Tr_{\sigma_a} \sigma_{a_1} \cdots
  \sigma_{a_r} g_n^z(\{\sigma_a \})}{Tr_{\sigma_a}
  g_n^z(\{\sigma_a \})} \ , 
 \end{equation}
we can write the eq. (\ref{EqG}) as:
\begin{equation} 
g_n(\{\sigma_a\}) =
 \cosh^n{(\beta J)} \sum_{r = 0}^{n} \left( \tanh^r{(\beta J)}
\sum_{a_1< \cdots <a_r} \sigma_{a_1}  \cdots \sigma_{a_r} 
q_{a_1 \cdots a_r} \right) \ .
\end{equation}
We can now implement the $\frac{1}{z}$ expansion if we scale  the couplings as usual:
\begin{equation} 
J = \frac{\tilde{J}}{\sqrt{z}} \ 
\end{equation}
and  set $\tilde{J} = 1$. 
Performing the expansion, after some computations we
obtain at the first order:
\begin{equation}
f= f_0 + \frac{1}{z} f_1 + O(\frac{1}{z^2}) \, ,
\end{equation}
with:
\begin{equation} \label{sk}
\beta f_0  =  -\frac{\beta^2}{4}+ \frac{\beta^2}{2n} \sum_{a<b} {q_{ab}^{(0)}}^2
-\frac{1}{n} \ln [Tr \exp( \beta \sum_{a<b} q_{ab}^{(0)}
 \sigma_a \sigma_b)] 
\end{equation}
and
\begin{eqnarray} \label{libpri}
\beta f_1 & = & - \frac{ \beta^2}{4}+ \frac{\beta^4}{24}- \frac{\beta^2}{2n}
 (1-\frac{5 \beta^2}{3}) \sum_{a<b} {q_{ab}^{(0)}}^2 -
\frac{\beta^4}{2n} \sum_{a<b<c<d} {q_{abcd}^{(0)}}^2 \\
& + & \frac{3 \beta^4}{n} \sum_{a<b<c} (q_{ab}^{(0)} q_{bc}^{(0)} q_{ca}^{(0)})
+ \frac{ \beta^4}{n} \sum_{a<b<c<d} (q_{ab}^{(0)} q_{cd}^{(0)} + 
q_{ac}^{(0)} q_{bd}^{(0)} + q_{ad}^{(0)} q_{bc}^{(0)}) \ q_{abcd}^{(0)} \ . \nonumber 
\end{eqnarray}
As it should be (we are expanding around $z= +\infty$),
$f_0$ is the SK free energy.
In these expressions we have also expanded the overlaps in powers of $1/z$:
\begin{eqnarray} \label{overlap} 
 q_{ab}  &= &
 q_{ab}^{(0)}  + \frac{1}{z}
 q_{ab}^{(1)}  + \cdots  \ , \nonumber\\
 q_{abcd}  &= &
 q_{abcd}^{(0)}  + \frac{1}{z}
 q_{abcd}^{(1)}  + \cdots  
\end{eqnarray}
and we have used the identities:
\begin{eqnarray}
q_{ab}^{(0)}&=&\frac{Tr_{\sigma} \exp{[\beta^2 \sum_{r<s} q_{rs}^{(0)}
\sigma_r \sigma_s]}
\sigma_a \sigma_b}{
 Tr_{\sigma} \exp{[\beta^2 \sum_{r<s} q_{rs}^{(0)}\sigma_r \sigma_s]}}
\equiv  <\sigma_a \sigma_b>_Q  \hspace{.3in} a\neq b \nonumber\\
&&\nonumber\\
 q_{abcd}^{(0)}&=&\frac{Tr_{\sigma} \exp{[\beta^2 \sum_{r<s} q_{rs}^{(0)}
\sigma_r \sigma_s]}
\sigma_a \sigma_b\sigma_c \sigma_d}{
 Tr_{\sigma} \exp{[\beta^2 \sum_{r<s} q_{rs}^{(0)}\sigma_r \sigma_s]}}
\equiv <\sigma_a \sigma_b\sigma_c \sigma_d>_Q
\hspace{.3in} a\neq b\neq c \neq d \, ,
\end{eqnarray}
where  $< \, \cdot \, >_Q$ is the average on  the 
single site Sherrington-Kirkpatrick Hamiltonian.
We notice that $f$ is no longer stationary with respect to the order parameter $ q$ 
because we have already used the stationary equations to simplify the result.

\subsubsection{Interaction with a Gaussian distribution.}
\noindent

If we use  a Gaussian distribution with the same mean ($\overline{J}=0$) 
and variance ($J_0^2/z$) of the previously studied bimodal distribution, one finds that at this 
order    the only relevant difference
 is in the fourth moment of the interaction
($3 J_0^4/z^2$ for the Gaussian and $J_0^4/z^2$ for the bimodal one),
 (it is crucial that at this order we expand the order parameter
$g_n$ only up to the second order in $z$, i.e. up to the fourth order
in $J$).
Performing the same calculations as before,
 we arrive to the final form for the 
free energy first order  correction:
\begin{eqnarray}\label{g} 
\beta f_1 & = & - \frac{ \beta^2}{4}+ \frac{\beta^4}{8}- \frac{\beta^2}{2n}
 (1- 3 \beta^2) \sum_{a<b} {q_{ab}^{(0)}}^2 -
\frac{3 \beta^4}{2n} \sum_{a<b<c<d} {q_{abcd}^{(0)}}^2 \\
& + & \frac{3 \beta^4}{n} \sum_{a<b<c} (q_{ab}^{(0)} q_{bc}^{(0)} q_{ca}^{(0)})
+ \frac{ \beta^4}{n} \sum_{a<b<c<d} (q_{ab}^{(0)} q_{cd}^{(0)} + 
q_{ac}^{(0)} q_{bd}^{(0)} + q_{ad}^{(0)} q_{bc}^{(0)}) q_{abcd}^{(0)} \ . 
\nonumber 
\end{eqnarray}
As expected, the $f_0$ does not change  because it does not contain $J^4$
terms and  the SK model is indeed independent from  the particular distribution
one uses if we fix the mean and the variance of the couplings.

\subsection{Random lattice with fluctuating connectivity.}
\noindent

In an other interesting model  the connectivity is a  
Poissonian variable with
mean value $z$. We take into consideration the large $z$
expansion, where the interactions probability distribution can be written
in the form:   
\begin{equation}
P(J_{ik})= (1- \frac{z}{N})\delta(J_{ik})+\frac{z}{N} \tilde{P}(J_{ik}) 
\hspace{.3in}
\forall i,k \hspace{.1in},
\end{equation}
where $\tilde{P}(J_{ik})$ is a distribution to be defined.

In principle we could write an expression similar to eq. (\ref{funz}) for the free energy, however 
it is simpler to proceed in a direct way.
The $n$ replicas partition function is:
\begin{eqnarray} \label{rand}
\overline{Z^n} &=& \prod_{i<k} \int_{-\infty}^{+\infty}P(J_{ik})dJ_{ik}
Tr_{\sigma}\exp(\beta J_{ik} \sum_{a}\sigma_i^a \sigma_k^a)\\
&=& \prod_{i<k}Tr_{\sigma} \left(1-\frac{z}{N} +\frac{z}{N}
\int_{-\infty}^{+\infty}\tilde{P(J_{ik})}dJ_{ik}
\exp(\beta J_{ik} \sum_{a}\sigma_i^a \sigma_k^a)\right) \nonumber \, .
\end{eqnarray}

\subsubsection{ The expression of the free energy}
\noindent

In the case where:
\begin{equation}
\tilde{P}(J_{ik})=  \frac{1}{2}[\delta(J_{ik}-J_0)+\delta(J_{ik}+J_0)]
 \hspace{.3in}
\forall i,k 
\end{equation}
we obtain:
\begin{eqnarray}
\overline{Z^n} &=& \prod_{i<k} Tr_{\sigma} \left(
1+ \frac{z}{N} 
[\cosh(\beta J_0 \sum_{a}\sigma_i^a \sigma_k^a) -1] \right)= \nonumber\\
&=& Tr_{\sigma} \exp\{\frac{z}{N} \sum_{i<k} 
[\cosh(\beta J_0 \sum_{a}\sigma_i^a \sigma_k^a) -1]\} \, .
\end{eqnarray}

Let us rescale:
  \begin{equation}
J_0 \rightarrow \frac{J_0}{\sqrt{z}} 
\end{equation}
and write  $J_0=1$. We can than perform the $1/z$ expansion up to
the first order for the free energy:
\begin{eqnarray}\label{svil}
\overline{Z^n} &= &Tr_{\sigma} 
\exp\{\frac{z}{N} \sum_{i<k} 
[ \frac{\beta^2}{2 z} {(\sum_{a}\sigma_i^a \sigma_k^a)}^2 +
\frac{\beta^4}{24 z^2} {(\sum_{a}\sigma_i^a \sigma_k^a)}^4 ]\} =
\\
&= &Tr_{\sigma} 
\exp\{\frac{1}{N} \sum_{i<k} 
[ \frac{\beta^2}{2 }
 {( \sum_{a,b} \sigma_i^a \sigma_k^a \sigma_i^b \sigma_k^b)} +
\frac{\beta^4}{24 z} {(\sum_{a,b,c,d} \sigma_i^a \sigma_k^a
\sigma_i^b \sigma_k^b \sigma_i^c \sigma_k^c \sigma_i^d \sigma_k^d)} ]\}
 \, .\nonumber
\end{eqnarray}
After converting the summations over  replicas indices   into 
distinct indices summations  (using $ {(\sigma^a)}^2 =1$),
introducing the Gaussian integrals and solving with the saddle point method,
 we find  at the first order:
  $\beta f = \beta f_0 + \frac{1}{z} \beta f_1$,
where $f_0$ is the SK free energy  and $f_1$ has the form:
\begin{equation}\label{ok}
\beta f_1 = \beta f_1^{J}\equiv  \frac{\beta^4}{24 }
+ \frac{\beta^4}{3 n}\sum_{a<b}{q_{ab}^{(0)}}^2
-\frac{\beta^4}{2 n}\sum_{a<b<c<d}{q_{abcd}^{(0)}}^2 \, .
\end{equation}

In a similar way, when $P(J)$ is a Gaussian distribution, we find the same result as before 
with the difference that the first order free energy $f_1$ is multiplied by a factor 
three.

 If we put together the previous  formulae we find that
 \begin{equation}\label{insieme} 
\beta f_1 = A f_{1}^{neigh} + K f_{1}^{J} \end {equation} where $A=1$ for the model with fixed 
number of neighbours, $A=0$ for the model with fluctuating number of neighbours and K is the kurtosis 
of the distribution of couplings $J$.  The quantity $f_{1}^{J}$ is given by eq.  (\ref{ok}) while 
$f_1^{neigh}$ by the formulae \ref{insieme} and \ref{g} is found to be:
\begin{eqnarray}\label{neigh} 
\beta f_1^{neigh} & = & - \frac{ \beta^2}{4}- \frac{\beta^2}{2n}
 (1-  \beta^2) \sum_{a<b} {q_{ab}^{(0)}}^2  \\
& + & \frac{3 \beta^4}{n} \sum_{a<b<c} (q_{ab}^{(0)} q_{bc}^{(0)} q_{ca}^{(0)})
+ \frac{ \beta^4}{n} \sum_{a<b<c<d} (q_{ab}^{(0)} q_{cd}^{(0)} + 
q_{ac}^{(0)} q_{bd}^{(0)} + q_{ad}^{(0)} q_{bc}^{(0)}) q_{abcd}^{(0)} \ . 
\nonumber 
\end{eqnarray}

\section{Evaluation of the sums over  replica's indices.}
\noindent

If  the replica is broken at two steps (using the usual conventions)
 we can write:
\begin{equation} \label{somuna}
\lim_{n \rightarrow 0} \frac{1}{n} \sum_{ab} q_{ab}^2 =
(m_2 -1) q_{2}^2 + (m_1 - m_2) q_{1}^2 + m_1 q_{0}^2  \ ,
\end{equation}
where the sum is on  the indices $a \neq b$. Similar expressions can be written at higher orders in 
the replica symmetry breaking and in the continuum limit one obtains:
\begin{equation} \label{somdue}
\lim_{n \rightarrow 0} \frac{1}{n} \sum_{ab} q_{ab}^2 =
\int_{0}^{1}  q^{2}(x).
\end{equation}
The formulae for the case where the replica symmetry is broken at  two steps 
can be obtained by using: \
\begin{eqnarray}
    q(x)=q_{0} \hspace{2cm}   0\leq x<m_{1}\\
      q(x)=q_{1} \hspace{2cm}  m_{1} \leq x<m_{2}\\
      q(x)=q_{2} \hspace{2cm}   m_{2}\leq x \leq 1
\end{eqnarray}  

The direct computation of 
\begin{equation}
\lim_{n \rightarrow 0} \frac{1}{n(n-1)(n-2)(n-3)} \sum_{abcd} 
 q_{abcd}^{2}
 \end{equation} 
 is more involved and if it is not properly done it can become a nightmare.

 \begin{figure}
\centerline{\psfig{figure=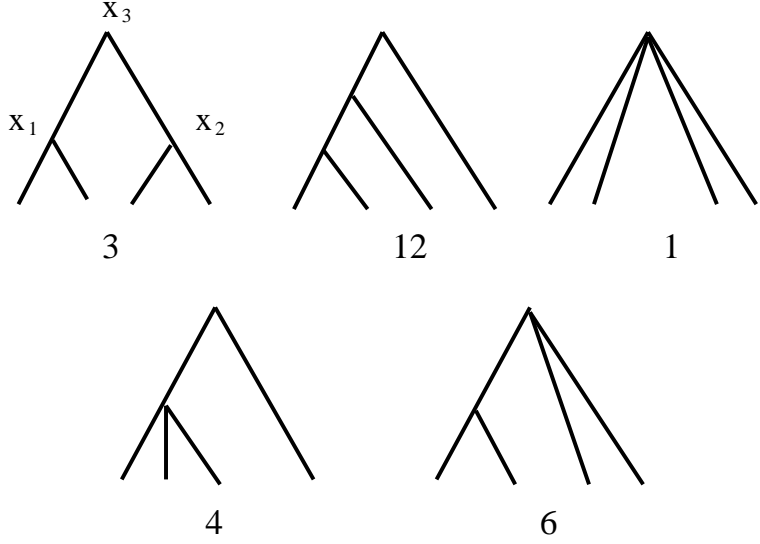,width=10cm}}
\caption[]{{ Four replicas diagrams.}
The number below the trees is the degeneration due to the replicas 
indices permutations.}
\label{figquattro}
\end{figure}

We can simplify it if we  remark that the four replicas overlap is  a function of all the possible
$x$'s among the  four replicas:
\begin{equation}
q_{abcd} = q(x_{12},x_{13},x_{14},x_{23},x_{24},x_{34})  ,
\end{equation}
with $0\leq x_{ij} \leq 1 \; \forall i, j$; however,
due to the ultrametric structure of the states at most
 three of the two replicas
 overlaps
 can be distinct. M\'ezard and Yedidia \cite{PS}  has shown that in order to compute this kind of 
 sums over replicas is suitable to 
consider the five possible ways in which four replicas 
can be organized (fig. \ref{figquattro}); we have to
associate a variable $x_i$ 
to each vertex and a factor  $x_i^{s-2} (s -2)!$ when $s$ lines converge
to it. Using this rule, providing to take into account the number of
different permutations of replicas indices that produce the same
 configuration, we finally obtain:
\begin{eqnarray} 
 &&\lim_{n \rightarrow 0} \frac{1}{n(n-1)(n-2)(n-3)} \sum_{abcd} 
 q_{abcd}^{2} =  3 \int_{1}^{0} dx_3 \int_{1}^{x_3}dx_2
\int_{1}^{x_3}dx_1  q(x_1, x_2, x_3)^{2}  + \nonumber\\
 &+&12 \int_{1}^{0} dx_3 \int_{1}^{x_3}dx_2
\int_{1}^{x_2}dx_1  q(x_1, x_2, x_3)^{2}  
+ \int_{1}^{0} dx_1 2 x_1^2 q(x_1^{2}) +\nonumber\\ 
&+& 4 \int_{1}^{0}dx_2
\int_{1}^{x_2} x_1 dx_1  q(x_1, x_2)^{2}  + 6 \int_{1}^{0} x_2 dx_2
\int_{1}^{x_2}  dx_1  q(x_1, x_2) ^{2} \hspace{0.3cm} .
\end{eqnarray}

\begin{figure}
\centerline{\psfig{figure=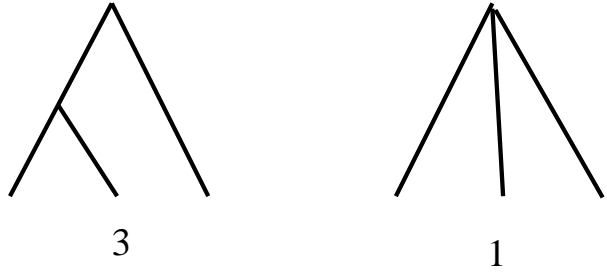,width=8cm}}
\caption[]{{ Three replicas diagrams.}  }
\label{figtre}
\end{figure}

At the second step of replica symmetry breaking we obtain:
\begin{eqnarray} \label{somqua}
  & \frac{24}{n} &\sum_{a<b<c<d}  q_{abcd}^{2}  = 
(m_2-1)(m_2-2)(m_2-3)q_{4_2}^2 \nonumber\\
            & + & 3(1-m_2)^2 (m_1-m_2) q_{2_2 2_2 sb}^2       
  -    3 m_1 (1-m_2)^2 q_{2_2 2_2 bd}^2 \nonumber\\
& - & 6 (1-m_2) (m_2-m_1) (2 m_2-m_1) q_{2_2 2_1 sb}^2 
        -  6 (1 - m_2) (m_2 - m_1) m_1 q_{2_2 2_1 bd}^2 \nonumber\\
  & - & 12(1 -m_2) m_1^2  q_{2_2 2_0}^2 
        +  (m_1-m_2)
             (m_1-2 m_2)(m_1-3 m_2) q_{4_1}^2 \nonumber\\
           &  - & 3(m_2-m_1)^2 m_1 q_{2_1 2_1}^2 
       -  12(m_2-m_1)
             m_1^2 q_{2_1 2_0}^2-6 m_1^3 q_{4_0}^2 \nonumber\\
      & + &       4(m_2-1) (m_2-2 )(m_1-m_2) q_{3_2 1_1}^2-
             4 m_1 (m_2-1) (m_2-2) q_{3_2 1_0}^2 \nonumber\\
      & - &       4 m_1 (m_1-m_2) (m_1-2 m_2) q_{3_1 1_0}^2-12
             m_1 (1-m_2)(m_2-m_1)q_{2_2 1_1 1_0}^2
\end{eqnarray}

      \begin{eqnarray} \label{quadue}
& \frac{24}{n} &\sum_{a<b<c<d}  q_{abcd}  (q_{ab} q_{cd} + q_{ac} q_{bd} + 
q_{ad} q_{bc}) = \nonumber\\
 & &   3 (m_2-1)(m_2-2)(m_2-3) q_{4_2} 
           q_2^2+3(1-m_2)^2(m_1-m_2)q_{2_2 2_2sb}
            (q_2^2+2q_1^2) \nonumber\\
& - & 4 m_1 (1-m_2)^2 q_{2_22_2bd}
             (q_2^2+2q_0^2)- 6(1-m_2)(m_2-m_1)
             (2m_2-m_1)q_{2_22_1sb}(q_2 q_1+  2 q_1^2) \nonumber\\
 &   - & 6(1-m_2)(m_2-m_1)m_1q_{2_22_1bd}
             (q_2q_1+2q_0^2)-12(1-m_2)m_1^2 q_{2_22_0}
             (q_2q_0+2q_0^2) \nonumber\\
& + & 3(m_1-m_2)(m_1-2m_2)
             (m_1-3m_2)q_{4_1}q_1^2-3(m_2-m_1)^2m_1
             q_{2_12_1}(q_1^2+2q_0^2) \nonumber\\
& - & 12(m_2-m_1)m_1^2
             q_{2_12_0}(q_1q_0+2q_0^2)-18 m_1^3 q_{4_0}
             q_0^2 \nonumber\\
& + & 12(m_2-1)(m_2-2)(m_1-m_2)
             q_{3_21_1}q_2 q_1-12 m_1(m_2-1)
             (m_2-2)q_{3_21_0}q_2
             q_0 \nonumber\\
& - & 12 m_1(m_1-m_2)(m_1-2m_2)q_{3_11_0}q_1q_0 \nonumber\\
          &   - & 12 m_1(1-m_2)(m_2-m_1)q_{2_21_11_0}(q_2q_0+2
             q_1q_0) \; , 
\end{eqnarray} 
where in the notation $q_{A_a \cdots}$ the quantity $A$ is the number of replicas and $a$ indicates 
the block to which they belong (we refer to the matrix $Q_{ab}$ at the second step of the 
ultrametric Ansatz); when two possibilities can occur, we write $sb$ when the four replicas are in 
the same block of first replica symmetry breaking and $bd$ in the other case (i.e. $q_{2_22_1sb}$ 
means that two replicas belong to the same second replica symmetry breaking block and all the four 
to the same block of first replica symmetry breaking; whereas $q_{2_22_1bd}$ means that two replicas 
belong again to the same second replica symmetry breaking block, the other two to the same first 
replica symmetry breaking block but the overlap between the first two and the second two is the 
minimum one).

For the sum on  three replicas indices we have (see fig.(\ref{figtre})):
\begin{eqnarray} \label{somtre}
& \frac{6}{n} & \sum_{a<b<c}
 (q_{ab}^{(0)} q_{bc}^{(0)} q_{ca}^{(0)}) =
          (m_2-1)(m_2-2)q_2^3 \nonumber\\
    & + &        3(1-m_2)(m_2-m_1)q_2q_1^2+
             3(1-m_2)m_1q_2q_0^2 \nonumber\\
    & + &         (m_1-m_2)(m_1-2m_2)q_1^3  +    
  3  m_1(m_2-m_1)q_1q_0^2+2m_1^2q_0^3 \ .
\end{eqnarray}
Substituting these expressions into  \ref{libpri},
\ref{g}, \ref{ok},
we obtain the explicit expression for  the free energies. 
We can now  find the numerical values of
 $q_2$, $q_1$, $q_0$, $m_2$, $m_1$ maximizing $f_0$ 
(the SK Hamiltonian is stationary with respect to  $q(x)$)
and then we can use  these values in the expressions of 
the four replicas overlaps.

To obtain the expressions at one level of RSB we can
 put $q_2=q_1$ and  identify the four replicas overlaps
in this way:
$q_{4_2}= q_{4_1}= q_{2_22_2sb}= q_{2_22_1sb}= q_{3_21_1}$;
$q_{3_21_0}= q_{3_11_0}= q_{2_21_11_0}$;
$q_{2_22_2bd}= q_{2_12_1}= q_{2_22_1bd}$;
$q_{2_22_0}= q_{2_12_0}$ (see \cite{mio}).

\section{The solution of the equation}
\noindent

To evaluate the  value of the free energy we have firstly to solve for the $m$ and $q$ 
parameters of the infinite connectivity limit and to compute the  parameters
 in equations (\ref{somdue}), (\ref{somqua}), (\ref{quadue}), (\ref{somtre}). At this end 
 we have to  compute integrals 
 like the following:

\begin{eqnarray}
q_{2_22_2sb}& =& \int_{-\infty}^{+\infty}  \frac{dz}{\sqrt{2 \pi q_0}}
\exp( - \frac{z^2}{2 q_0}) 
\left[\frac{\mbox{Num}}{\mbox{Den}}\right] \nonumber\\
&&\mbox{where} \nonumber\\
\mbox{Num}& =& 
 \int_{-\infty}^{+\infty} dy \exp( - \frac{y^2}{2 (q_1 - q_0)}) 
 \left\{ \int_{-\infty}^{+\infty} dx
 \exp( - \frac{x^2}{2 (q_2 - q_1)})  \cosh^{m_2}(\beta (z+y+x))
\begin{array}{c} \\ \\ \end{array}
 \right\}^{ \frac{m_1}{m_2} - 2}
 \nonumber\\
& \times &
\left\{ \int_{-\infty}^{+\infty} dx \exp( - \frac{x^2}{2 (q_2 - q_1)}) 
 \tanh^2(\beta (z+y+x))\right.  \cosh^{m_2}(\beta (z+y+x)) \left.
 \begin{array}{c} \\ \\ \end{array}\right\}^2
  \nonumber\\
&& \mbox{and} \nonumber\\
\mbox{Den} &=&  
 \int_{-\infty}^{+\infty} dy \exp( - \frac{y^2}{2 (q_1 - q_0)}) 
 \left\{ \int_{-\infty}^{+\infty} dx 
\exp( - \frac{x^2}{2 (q_2 - q_1)}) 
\cosh^{m_2}(\beta (z+y+x)) \right\}^{\frac{m_1}{m_2}} \ . 
 \end{eqnarray}

To evaluate these expressions it is important to optimize the number of operations the computer has 
to do.  In the numerical evaluation of the integrals (we are considering sums instead of integrals 
and we set $x=a \  i$ where $i$ is an integer) the internal integrals have in fact to be evaluated for 
every value of the variable of the external one.  A repeated evaluation would take an enormous 
amount of time.  A much faster method consists in evaluating beforehand the internal functions (i.e. 
$\cosh(x+y+z)$) for all values $x+y+z=a \ i$ and in storing in a table the values in the integrals; in 
this way the computer has to perform a number of operations proportional to $N$ instead to $N^3$ of 
the naive method
($N$ is the number of spacings in which the integration domain is divided).
\begin{figure}
\begin{tabular}{cc}
\includegraphics[width=8.5cm]{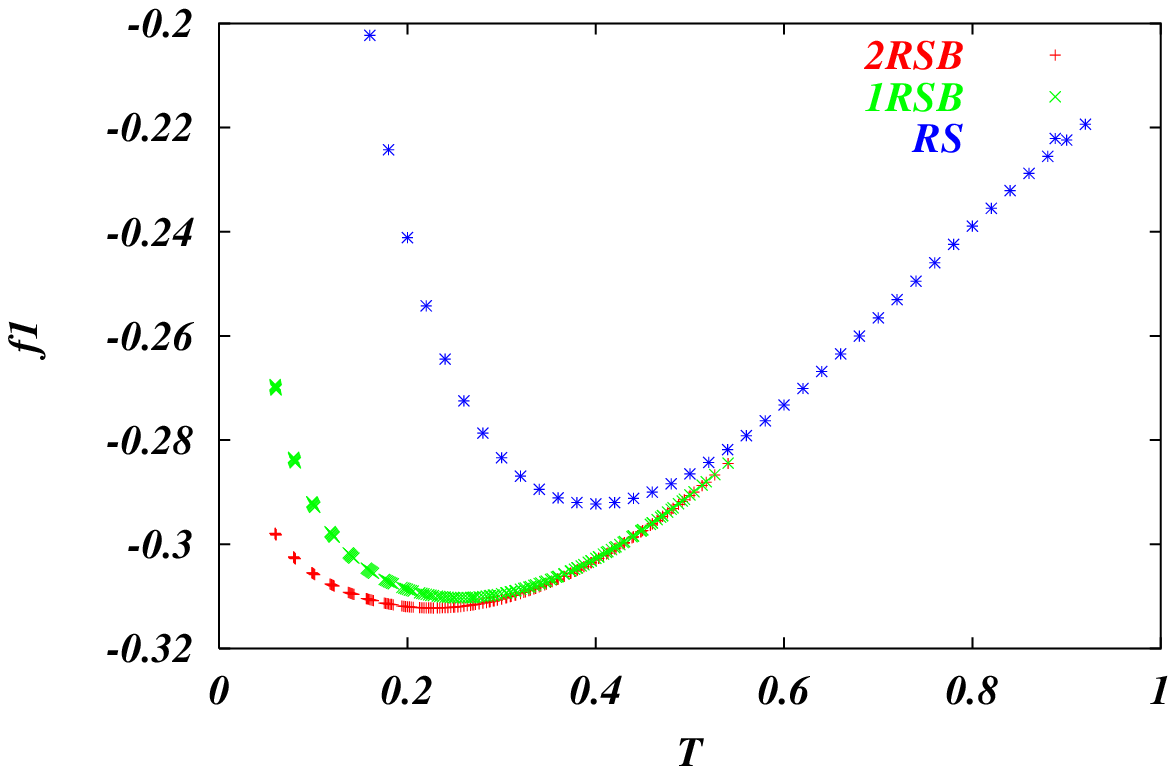}&
\includegraphics[width=8.5cm]{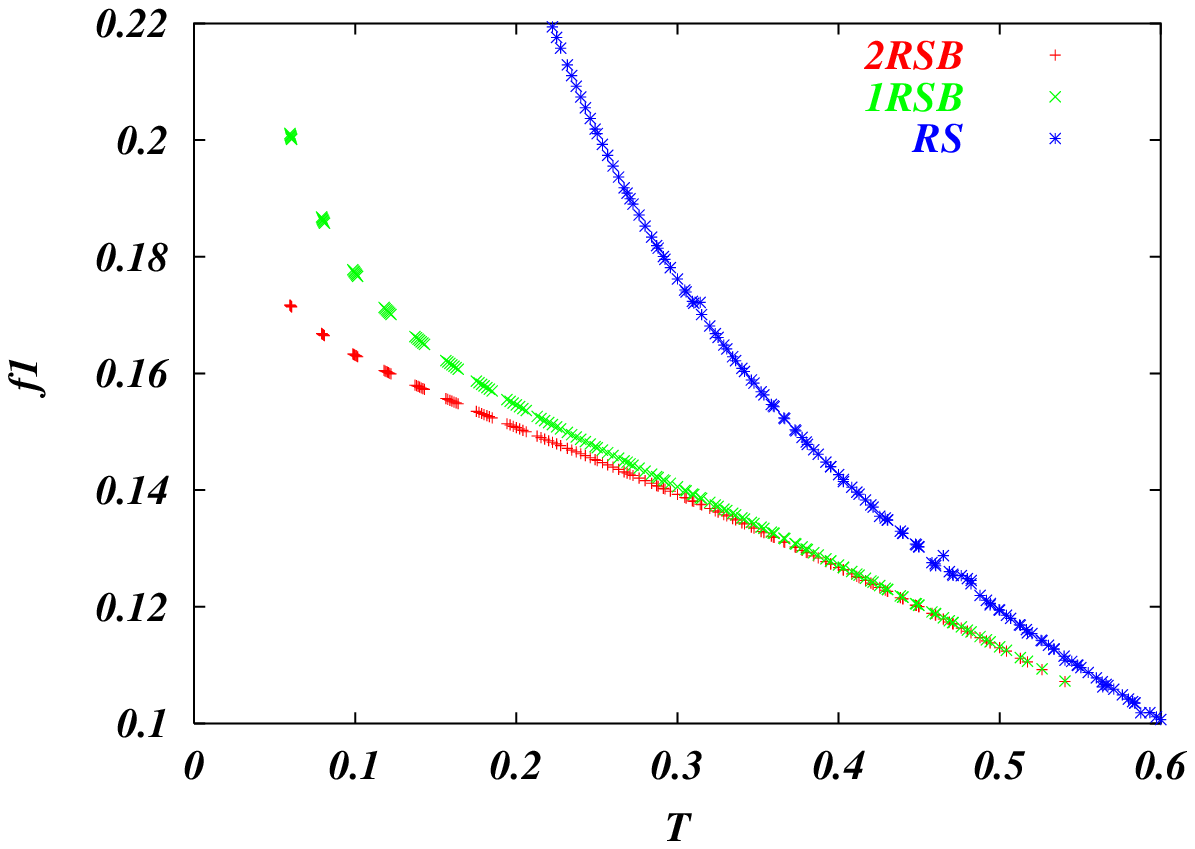}
\end{tabular}
\caption[]{{ The $1/z$ correction to the free energy as function of the temperature for the replica 
symmetric case ($*$), one step replica symmetry breaking and two steps replica symmetry breaking 
for the model with $J=\pm 1$ for fixed connectivity (left) and for fluctuating connectivity (right),}  }
\label{figfree}
\end{figure}

\begin{figure} \begin{center}
\includegraphics[width=10.5cm]{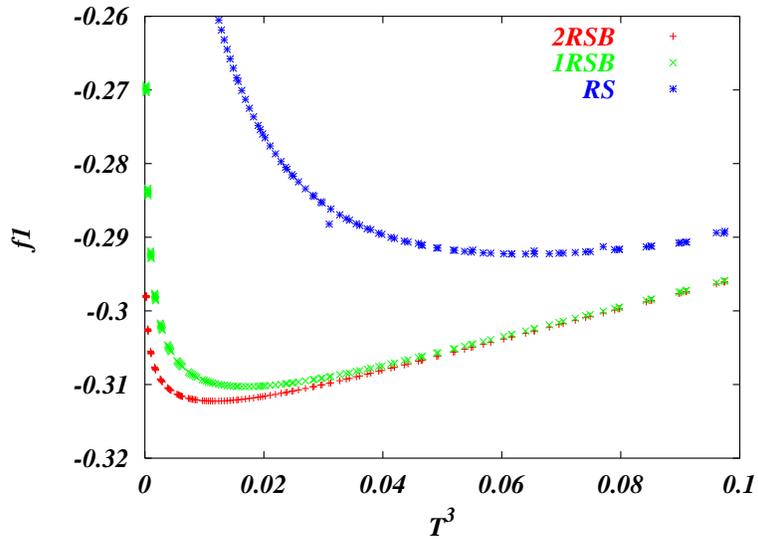}
\end{center}
\caption[]{{ The $1/z$ correction to the free energy as function of $T^3$ for the replica 
symmetric case ($*$), one step replica symmetry breaking and two steps replica symmetry breaking 
for the model with $J=\pm 1$ for fixed connectivity }  }
\label{Ttre}
\end{figure}

The final results for the coefficient of the $1/z$ corrections to the free energy are shown in fig 
\ref{figfree} in the case of $\pm 1$ interactions. We immediately see that the divergence of the 
correction to the free energy  at  $T=0$ fades away when we increase  the order of the replica 
breaking  and it is an artefact of using  a starting point which is not correct (the correct one 
corresponds to infinite breaking of the replica symmetry).

\begin{figure} \begin{center}
\includegraphics[width=10.5cm]{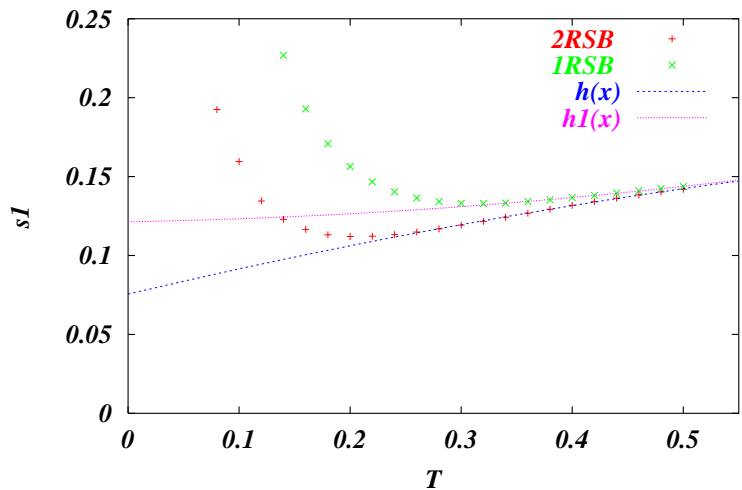}
\end{center}
\caption[]{{ The $1/z$ correction to the entropy as function of the temperature for one step replica symmetry breaking and two steps replica symmetry breaking 
for the model with $J=\pm 1$ for fixed connectivity and the corresponding second order polynomial 
fits in the high temperature region.}  }
\label{figentro}
\end{figure}

We evaluated the entropy doing the derivative of the free energy ($S=-df/dT)$ using an high order 
expression for the finite difference derivative.  The final results for the coefficient of the $1/z$ 
corrections to the free energy are shown in fig \ref{figentro} in the case of $\pm 1$ interactions 
with fluctuating connectivity.  Also in this case we see that the divergence of the correction to 
the entropy near $T=0$ fades away when we increase the order of the replica breaking. The 
correction for the entropy are much stronger that those for the free energy. In order to evidentiate 
the effect of the spurious divergence at $T=0$ we show also a second order polynomial fit in the 
high temperature region, which dramatically fails at low temperature.

\section{Conclusions.}
\noindent

We can see from the numerical data that the first order correction $f_1$
of the free energy in  all the analyzed  models taken into account presents
a divergence at small temperatures. We fit in the interval
$T \in [0.05,0.5]$ a behaviour of the kind (see
fig.(\ref{figfree})):
\begin{equation}
f_1(T)= D + A T^3 + B T^2 + \frac{C}{T} \; .
\end{equation} 
We report in the following  table the values of the $C$ coefficient  in
the different cases considered.\\
\vspace{0.5cm}\\
\begin{center}
\begin{tabular}{|l|c|c|c|c|}
\hline
\multicolumn{5}{|c|}{$C$ values}\\
\hline
&\multicolumn{2}{c|}{fixed connectivity}&
\multicolumn{2}{c|}{fluct. connectivity}\\
\cline{2-5}
& $\pm 1$ &  Gaussian&  $\pm 1$&  Gaussian\\
\hline
RS&0.035&0.12&0.35&0.12\\
\hline
1RSB&0.003&0.010&0.003&0.010 \\
\hline
2RSB&0.001&0.003&0.001&0.003 \\
\hline
\end{tabular}
\end{center}
\vspace{0.5cm}

This coefficient is three times smaller when going from the 1RSB solution to the 2RSB 
one.  Moreover, if we look at the results in \cite{ultimo} we can see that the same ratios have been 
found for the $1/\sqrt{z}$ coefficient; its values are in fact $0.1$ for RS, $0.01$ for 1RSB and 
$0.0026$ for 2RSB solutions.  The same divergent factors occur in the fixed and in the fluctuating 
connectivity models and the entire divergence comes from the term  proportional the forth moment of 
the distribution of the $J$.
However, we can see from the figures that the divergence moves to smaller temperatures when the 
number of replica symmetry breaking steps increases.  In the fixed connectivity model with bimodal 
distribution the divergence appears at $T < 0.1$.

We notice that in all these models there is a $T^2$ correction to the low temperature behaviour
of the S.K. model, where the free energy is proportional to $T^3$.
From the numerical data suggest that this correction goes to zero in the full
replica symmetry breaking solution, at least  in the 
fixed connectivity model, that is linear in $T^3$ over
a large range of temperature already in the 2RSB solution (fig.(\ref{Ttre})).

We fitted the  curves in the temperature interval where the divergence doesn't yet occur,
with the function:
\begin{equation}
f_1(T)= D + A T^3 + B T^2  \; .
\end{equation} 
 For completeness we  report in the following  table the values of the  coefficients $A$ and $B$ in
the different cases considered.\\
\vspace{0.5cm}\\
\begin{center}
\begin{tabular}{|l|c|c|c|c|}
\hline
\multicolumn{5}{|c|}{$B$ values}\\
\hline
&\multicolumn{2}{c|}{fixed connectivity}&
\multicolumn{2}{c|}{fluct. connectivity}\\
\cline{2-5}
& $\pm 1$ &  Gaussian&  $\pm 1$&  Gaussian\\
\hline
RS&-0.199&-0.049&-0.63&-1.88\\
\hline
1RSB&-0.122&-0.139&-0.37&-1.12 \\
\hline
2RSB&-0.054&-0.08&-0.3&-0.9\\
\hline
\end{tabular}
\end{center}
\vspace{0.5cm}
\begin{center}
\begin{tabular}{|l|c|c|c|c|}
\hline
\multicolumn{5}{|c|}{$A$ values}\\
\hline
&\multicolumn{2}{c|}{fixed connectivity}&
\multicolumn{2}{c|}{fluct. connectivity}\\
\cline{2-5}
& $\pm 1$ &  Gaussian&  $\pm 1$&  Gaussian\\
\hline
RS&-0.286&-0.156&0.19&0.619\\
\hline
1RSB&-0.30&-0.176&0.165&0.497 \\
\hline
2RSB&-0.313&-0.18&0.16&0.48 \\
\hline
\end{tabular}
\end{center}
\vspace{0.5cm}

The fit we have done assumes that, apart form the $1/T^{2}$ divergence, the entropy extrapolates to 
zero at zero temperature\footnote{This is true for the Gaussian model, but it is not true for the 
$\pm 1$ at fixed $z$, where {\sl spin fou} are present.  However it is reasonable that this 
difference can be seen only at higher orders in the $1/z$ expansion.  }.  In order to check the 
consistency of the results, we have extrapolated to zero temperature the numerical results for the 
entropy in the temperature interval where the divergence doesn't appear yet.  We then find a 
behaviour that accords with the expected one: the zero temperature value $s(0)$ is different from 
zero but decreases when going to higher steps of replica symmetry breaking, suggesting that it 
will reach the correct value $s_1(0)=0$ in the infinite steps limit.

We give the results in the following table.  \\
\vspace{0.5cm}\\
\begin{center}
\begin{tabular}{|l|c|c|c|c|}
\hline
\multicolumn{5}{|c|}{1st order correction to entropy at $T=0$}\\
\hline
&\multicolumn{2}{c|}{fixed connectivity}&
\multicolumn{2}{c|}{fluct. connectivity}\\
\cline{2-5}
 & $\pm 1$ &  Gaussian&  $\pm 1$&  Gaussian\\
\hline
1RSB&0.26&0.48&0.12&0.36 \\
\hline
2RSB&0.18&0.32&0.07&0.22 \\
\hline
\end{tabular}
\end{center}
\vspace{0.5cm}

To conclude, we think that there are numerical evidences that confirm
 that the  $1/z$ expansion is correct  to study the random lattices
with  high connectivity. 
The  $1/z$ expansion  arise naturally if we compare
the  high  connectivity limit  in random lattices 
to the  case of nearest neighbor interactions in the high dimension limit.
The high dimension expansion is indeed in powers of $1/D$ and it coincides at the first order in $D$ 
with the $1/z$ expansion.

\section{Appendix.}
\noindent
In order to demonstrate that (\ref{funz}) is the correct functional for
the free energy
we can show that $\lim_{n \rightarrow 0} \frac{1}{n}
 \partial (\beta f)/ \partial \beta$  
is the internal energy when $g_n$ is solution of
 (\ref{general}) and that (\ref{funz})   is
correctly normalized at $\beta = 0$
($- \beta f(\beta=0) = \ln 2$); the latter condition is easy to
verify
considering that $g_n(\beta = 0) = 1$ and that
 $Tr_{\sigma_a}$
gives $2^n$ terms. To convince oneself of the validity of the former
assertion one can construct explicitly the order parameter
$g_{n}(\{\sigma_0^a\})$ making clear its physical meaning.
Following the approach of \cite{cavita} we start writing  the partition function  in a recursive manner
 making use of the equivalence of the model
with  a Cayley tree. Focusing on an  arbitrary
spin $\sigma_0$:
 \begin{equation}
Z =  \sum_{\{\sigma\}} \exp{(\beta h \sigma_0)}
\prod_{k = 1}^{z + 1} \mathit{Q_{(L)}}
(\sigma_0|\sigma^{(k)}) \ ,
\end{equation}
where
\begin{equation}
\mathit{Q_{(L)}} (\sigma_0|\sigma^{(1)}) 
= \exp{(\beta J_{01} \sigma_0 \sigma_1 + \beta h \sigma_1)}
\prod_{k = 1}^{z} \mathit{Q_{(L - 1)}} (\sigma_1|\sigma^{(k)}) \ .
\end{equation}
 $z + 1$ is the  branches number (random lattice's connectivity)
 and $L$ the shells number; the $\sigma^{(k)}$
are the spins on $k$-th branch  excluding the 
 $\sigma_0$; $h$ is an external uniform field.

We can than write the $n$ replicas partition function:
\begin{equation}\label{uno}
Z^n =   \prod_{a = 1}^{n} Z_a =
\sum_{\{ \sigma_1 \}} \cdots \sum_{\{ \sigma_n \}} 
\exp{\left(\sum_{a = 1}^{n} \beta h \sigma_0^a \right)}
\prod_{k = 1}^{z + 1} \prod_{a = 1}^{n}\mathit{Q_{(L)}}
(\sigma_0^a|{\sigma^{(k)}}^a) 
\end{equation}
and define:
\begin{equation}\label{due}
g_{n,(L)}(\{\sigma_0^a\}) \equiv \overline{\sum_{\{ {\sigma^{(k)}}^a \}}
\prod_{a = 1}^{n}\mathit{Q_{(L)}}
(\sigma_0^a|{\sigma^{(k)}}^a)} \hspace{0.5cm} ,
\end{equation}
where the bar is the average over the random couplings $J$.

From (\ref{uno}) and (\ref{due}) we obtain:
\begin{equation} \label{rami}
\overline{Z^n} = \sum_{\{ \sigma_0^a \}}
\exp{\left(\sum_{a = 1}^{n} \beta h \sigma_0^a \right)}
g_{n,(L)}^{z + 1}(\{\sigma_0^a\}) \hspace{0.5cm} ,
\end{equation}
that reveals $ g_{n,(L)}(\{\sigma_0^a\})$ to be the one branch contribution to
the partition function.

By definition it follows the recursion relation:
\begin{equation}\label{recursion}
g_{n,(L)}(\{\sigma_0^a\}) = \int_{-\infty}^{+\infty}dJ P(J)
\sum_{\{ \sigma_1^a \}} 
\exp{\left( \sum_{a = 1}^{n} \beta h \sigma_1^a +
\sum_{a = 1}^{n} \beta J \sigma_0^a \sigma_1^a \right)}
g_{n,(L - 1)}^z (\{\sigma_1^a\}) \ .
\end{equation}

The  internal energy density can be written as a bond energy
 multiplied by the number of links per spin ($(z+1)/2$). 

If we consider a link with a coupling constant $J$
between two spins $\sigma_0$ and
$\sigma_1$, we can write its energy as \cite{cavita}:
\begin{equation}
E_{01}=-J <\sigma_0 \sigma_1> .
\end{equation}
The expectation value is computed with the Hamiltonian
$H= - J\sigma_0 \sigma_1 + H_0 +H_1$, where $H_0$ is the
Hamiltonian of the spin $\sigma_0$
before being connected with $\sigma_1$ and can be written as
$H_0 = -\ln \left(g_{n}(\{\sigma_0\})\right)^z/\beta$;
the same argument can be repeated for $\sigma_1$.

At this level we should use the finite normalized order parameter:
\begin{equation}
g_{n,(L)}(\{\sigma_0^a\}) \equiv \frac{ \overline{\sum_{\{ {\sigma^{(k)}}^a \}}
\prod_{a = 1}^{n}\mathit{Q_{(L)}}
(\sigma_0^a|{\sigma^{(k)}}^a)} }{\overline{\sum_{\{ {\sigma_k}^a \}}
 \prod_{l = 1}^{z}\sum_{\{ {\sigma^{(l)}}^a \}}
\prod_{a = 1}^{n}\mathit{Q_{(L - 1)}}
(\sigma_k^a|{\sigma^{(l)}}^a)} } \ 
\end{equation}
which follows the recursion equation:
\begin{equation} 
g_{n,(L)}(\{\sigma_0^a\}) = \frac{ \int_{-\infty}^{+\infty}dJ P(J)
\sum_{\{ \sigma_1^a \}} 
\exp{\left( 
\sum_{a = 1}^{n} \beta J \sigma_0^a \sigma_1^a \right)}
g_{n,(L - 1)}^z (\{\sigma_1^a\})}{
\sum_{\{ \sigma_1^a \}}g_{n,(L - 1)}^{z }(\{\sigma_1^a\})} \ . 
\end{equation}
In the thermodynamic limit, taking into account only the inner part of 
the Cayley tree,
the  $g_n$ are shells independent, so
we can write:
\begin{equation} \label{normal}
g_n(\{\sigma_a\}) = \frac{ \int_{-\infty}^{+\infty}dJ P(J)
\sum_{\{ \tau_a \}} 
\exp{\left( 
\sum_{a = 1}^{n} \beta J \sigma_a \tau_a \right)}
g_n^z (\{\tau_a\})}{
\sum_{\{ \tau_a \}}g_n^{z }(\{\tau_a\})}  \ .
\end{equation}
 
Anyway the internal energy (as the free energy)
 is insensitive to the normalization of the order parameter,
so we can use both (\ref{normal}) or (\ref{recursion})
(in the last one the thermodynamic limit has to be taken).

We can now evaluate the derivative of $\beta f$ with respect
to $\beta$ and appurate that we obtain the expression we
expected:  
\begin{eqnarray} \label{interna}
&&n\frac{\partial(\beta f)}{\partial \beta} =
\frac{z + 1}{2} \times \\
&&\times \frac{\int_{-\infty}^{+\infty}dJ P(J)
Tr_{\{\sigma_a\}}Tr_{\{\tau_a\}} g_n^{z }(\{\sigma_a\})
 g_n^{z }(\{\tau_a\}) \exp{[\beta J \sum_{a = 1}^{n}\sigma_a
\tau_a]}(- J \sum_{a = 1}^{n}\sigma_a
\tau_a)}
{ \int_{-\infty}^{+\infty}dJ P(J)
Tr_{\{\sigma_a\}}Tr_{\{\tau_a\}} g_n^{z }(\{\sigma_a\})
 g_n^{z }(\{\tau_a\}) \exp{[\beta J \sum_{a = 1}^{n}\sigma_a
\tau_a]}} \nonumber .
\end{eqnarray}

\end{document}